\documentclass[conference]{IEEEtran}
\usepackage[table,xcdraw]{xcolor}
\usepackage{url}
\usepackage{latexsym}
\usepackage{graphicx}
\usepackage{makecell}
\usepackage{array}
\usepackage{hyperref}
\usepackage{bbm}
\usepackage{tikz}
\usepackage{amsmath}
\usepackage[numbers]{natbib}
\usepackage{booktabs}
\usepackage{fourier}

\title{ACES: Evaluating Automated Audio Captioning Models on the Semantics of Sounds}

\author{%
\IEEEauthorblockN{%
Gijs Wijngaard\IEEEauthorrefmark{1}, Elia Formisano\IEEEauthorrefmark{2}, Bruno L. Giordano\IEEEauthorrefmark{3} and Michel Dumontier\IEEEauthorrefmark{1}%
}%
\IEEEauthorblockA{\IEEEauthorrefmark{1} 
Department of Advanced Computing Sciences,
Faculty of Science and Engineering,
Maastricht University \\
}%
\IEEEauthorblockA{\IEEEauthorrefmark{2} 
Department of Cognitive Neuroscience,
Faculty of Psychology and Neuroscience,
Maastricht Centre for Systems Biology \\ 
Faculty of Science and Engineering,
Brightlands Institute for Smart Society,
Maastricht University \\
}%
\IEEEauthorblockA{\IEEEauthorrefmark{3} 
Institut des Neurosciences de La Timone, UMR 7289, CNRS and Université Aix-Marseille}
}

\date{}

\begin{document}

\maketitle
\begin{abstract}

Automated Audio Captioning is a multimodal task that aims to convert audio content into natural language. The assessment of audio captioning systems is typically based on quantitative metrics applied to text data. Previous studies have employed metrics derived from machine translation and image captioning to evaluate the quality of generated audio captions. Drawing inspiration from auditory cognitive neuroscience research, we introduce a novel metric approach -- Audio Captioning Evaluation on Semantics of Sound (ACES). ACES takes into account how human listeners parse semantic information from sounds, providing a novel and comprehensive evaluation perspective for automated audio captioning systems. ACES combines semantic similarities and semantic entity labeling. ACES outperforms similar automated audio captioning metrics on the Clotho-Eval FENSE benchmark in two evaluation categories.
\end{abstract}
\begin{IEEEkeywords}
automated audio captioning, evaluation metric, sound semantics	
\end{IEEEkeywords}

\section{Introduction}
Automated audio captioning (AAC) is an emerging field of audio processing. As introduced in 2017 by \citet{drossosAutomatedAudioCaptioning2017}, the goal of AAC is to describe the content of an audio clip using natural language, that is, using structured text that describes sound. The performance of AAC models is estimated using metrics that compare model-predicted captions with human annotations.

Standard AAC models are based on the encoder-decoder architecture \cite{meiAutomatedAudioCaptioning2022}, where the encoder converts the input audio to an embedding, which is then fed to the decoder. In this architecture, the decoder learns to generate the audio caption by minimizing the cross-entropy loss on the probabilities of the decoder outputs. During inference, the decoder calculates the most probable sentence given the embedded audio.

AAC models are benchmarked using metrics tailored to various criteria (see Section \ref{sec:drawbacks}, below). Additionally, it is possible to benchmark the metrics themselves, and gauge their alignment with human judgment. In this context,  \citet{zhouCanAudioCaptions2022} introduced the FENSE benchmark, which evaluates how AAC metrics perform in comparison to human evaluation. 

Research in the field of auditory cognition and neuroscience has shown that humans listen to sounds to derive information on sources, events, and changes in the environment and that this information is reflected in listeners' verbal descriptions of everyday sounds \cite{giordanoWhatWeMean2022a}. When asked to describe sounds, listeners refer to the presence of animate (\textit{who}) or inanimate (\textit{what}) sources, identify mechanisms or actions of sound generation (\textit{how}) and, eventually, to a spatial (\textit{where}) or temporal (\textit{when}) context (see Table \ref{tab:lab-cat}, and \cite{giordanoWhatWeMean2022a}, for the derivation of these classes of descriptors from previously published taxonomies of natural sounds).

In this paper, we propose a novel metric to evaluate audio captioning algorithms on annotated datasets: \textbf{A}udio \textbf{C}aptioning \textbf{E}valuation on Semantics of \textbf{S}ounds'' (ACES)\footnote{Code, data and models available at: \href{https://github.com/GlJS/ACES}{https://github.com/GlJS/ACES}.}. ACES draws inspiration from auditory cognition research on the extraction of sound descriptors: words or phrases that describe meanings such as what makes the sound, what is the location and how the sound is produced (see Table \ref{tab:lab-cat} for a list of all categories). This is related to semantic role labeling, which is the process of assigning labels to words to indicate its semantic role \cite{boas2017past}. ACES utilises the cosine similarity of vector representations of these sound descriptors to measure the correspondence between generated and ``ground truth'' captions. ACES returns a score that indicates the quality of the generated caption based on these ground truths.

In short, the ACES score consists of a cosine similarity of the pairwise token combination. ACES uses precision and recall from this cosine similarity, and takes the token with the highest score as overlap. ACES also utilises a METEOR based approach in weighing of precision and recall, a small penalty for longer captions, and a fluency error detection model. This work is an extension of an earlier version of ACES, presented at European Signal Processing Conference 2023 (EUSIPCO) \cite{wijngaard2023aces}. The current version of ACES supersedes the initial release, rendering the previous version obsolete and deprecated.

\section{Related work}

The evaluation metrics from the Microsoft COCO Caption Dataset represent a tool-set of audio captioning evaluation algorithms \cite{chenMicrosoftCOCOCaptions2015}. This set of evaluation metrics was originally intended to measure performance in image captioning tasks, but has since been adopted in other domains, including audio captioning. The tool-set includes the metrics BLEU \cite{papineniBleuMethodAutomatic2002}, ROUGE \cite{linROUGEPackageAutomatic2004}, METEOR \cite{banerjeeMETEORAutomaticMetric2005} and CIDEr \cite{vedantamCIDErConsensusBasedImage2015}, and, in later COCO versions, SPICE \cite{andersonSPICESemanticPropositional2016}. The current standard in AAC metrics are SPIDEr and its derivative SPIDEr-FL. SPIDEr is a combination of CIDER and SPICE, which outperforms both metrics based on human evaluation of a randomly sampled COCO test set (10.56\% increase compared to a baseline MLE model) \cite{liuImprovedImageCaptioning2017}. SPIDEr-FL is SPIDEr enhanced with an additional pretrained fluency detection model, specifically designed to identify incomplete sentences and sentences with repeated or missing words \cite{zhouCanAudioCaptions2022}. For an overview of recent metrics for evaluating Automated Audio Captioning models, see Table \ref{tab:overview-metrics}.

\begin{table}[]
\caption{Overview of metrics used for evaluating Automated Audio Captioning models}
\label{tab:overview-metrics}
\begin{tabular}{@{}lll@{}}
\toprule
Metric     & Description                                                                                                                                                                                    & Primary use         \\ \midrule
BLEU$_{1..4}$ & \begin{tabular}[c]{@{}l@{}}Uses the co-occurrences of n-grams. \\ Adds a brevity penalty.\end{tabular}                                                                                         & Machine translation \\ \midrule
METEOR     & \begin{tabular}[c]{@{}l@{}}Harmonic mean of precision and \\ recall of caption chunks. Recall \\ weighted 9 times more than \\ precision. Matches also stems, \\ synonyms and paraphrases\end{tabular} & Machine translation \\ \midrule
ROUGE      & \begin{tabular}[c]{@{}l@{}}Calculates F-measure using,\\ longest common subsequence.\\ Oriented to recall with $\beta = 1.2$.\end{tabular}                                                     & Text summarization  \\ \midrule
SPICE      & \begin{tabular}[c]{@{}l@{}}Matches captions on object classes,\\ relation types and attributes.\end{tabular}                                                                                   & Image captioning    \\ \midrule
CIDER      & \begin{tabular}[c]{@{}l@{}}Calculates a weighted sum of the \\ cosine similarity for n-grams \\ with $n \in [1,4]$. Uses TF-IDF.\end{tabular}                                                  & Image captioning    \\ \midrule
BertScore  & \begin{tabular}[c]{@{}l@{}}Calculate F1, precision and recall \\ of matched words from the \\ pairwise cosine similarity \\ of the sentence embeddings.\end{tabular}                           & Text generation     \\ \midrule
BLEURT     & \begin{tabular}[c]{@{}l@{}}BERT model trained on human \\ annotated dataset of sentence pairs,\end{tabular}                                                                                    & Text generation     \\ \midrule
FENSE      & \begin{tabular}[c]{@{}l@{}}Cosine similarity of sentence \\ embeddings combined with \\ a fluency detection model\end{tabular}                                                                 & Audio captioning    \\ \midrule
SPICE+     & \begin{tabular}[c]{@{}l@{}}Combines SPICE with matching \\ using sentence embeddings\end{tabular}                                                                                              & Audio captioning    \\ \bottomrule
\end{tabular}
\end{table}

\subsection{Drawbacks of current metrics} \label{sec:drawbacks}

The performance metrics currently adopted in the field of AAC are characterized by several drawbacks. For example, the BLEU, ROUGE, CIDEr and METEOR metrics are sensitive to n-gram overlap, which is neither necessary nor sufficient for establishing meaning similarity between sentences \cite{andersonSPICESemanticPropositional2016}. For example, the captions ``Rain coming from a big cloud'', and ``Music coming from a big band'', are dissimilar in meaning but have a 4-gram in common, resulting in a high n-gram score. Although SPICE was proposed to mitigate this problem, it makes the overly stringent assumption that the compared captions have exactly the same wording for its entities, attributes, and relations. For example, the candidate sentence ``Young woman talking with crunching noise'' and the reference sentence ``Paper crackling with female speaking lightly in the background.'' result in a SPICE score of 0, despite their clear affinity in meaning \cite{zhouCanAudioCaptions2022}. In other cases, the caption may not include text descriptors of the sound-generating agents or objects (e.g. ``Metallic scraping that stops and then starts again'') or descriptors of actions (e.g. ``Very loud static sound without any other noise''). In such cases, a scene graph (a way of encoding objects, attributes and relations) cannot be composed or calculated, and the SPICE score returns a low value regardless of semantic relatedness.

Here, a novel metric is proposed based on semantic role structure (e.g. who did what to whom). Semantic-role was previously considered by \citet{loFullyAutomaticSemantic2012}, who proposed the MEANT metric in the context of Machine Translation. A drawback of this metric is the use of syntactic parsing of the sentence, which results in a low score when synonyms are used. Pretrained models such as BLEURT \cite{sellamBLEURTLearningRobust2020} and BERTScore \cite{zhangBERTScoreEvaluatingText2020} have also been used as a metric for various machine learning tasks. One of the benefits of these models is that you can extract a contextualized embedding that captures the meaning of the term, and that similar meanings are correlated with the computational similarity derived from a cosine of the two embeddings.

The comparison of sentences based on lexical structure has been investigated in recent studies on Automated Audio Captioning. The FENSE metric \cite{zhouCanAudioCaptions2022} uses a language model to calculate similarity and includes a fluency penalty model to capture coherent structures in audio captions. Specifically, FENSE penalizes captions in five categories: incomplete sentences, repeated events, repeated adverbs, missing conjunctions, and missing verbs. This fluency penalty was also added to the SPIDEr metric in the DCASE Challenge 2023 Task 6a \cite{stowellDetectionClassificationAcoustic2015a}. The SPICE+ metric \cite{gontierSPICEEvaluationAutomatic2022} finally tries to solve issues that arise with its counterpart, SPICE, by carrying out the evaluation within a language model framework. 

\subsection{Research in other fields}

Analogies can be drawn between our research and active areas within natural language processing (NLP), specifically: (i) semantic role labeling (SRL), where predicate-argument structures are modeled from sentences, and (ii) part-of-speech tagging (POS), where words are tagged based on their grammatical functions (e.g. subject or verb). While our dataset is directly labeled by us, there is a clear relation between our labels and the classifications that an automated SRL or POS model would assign. For example, the \textit{ARG-0} and \textit{V} labels in a SRL model correspond, respectively, to the WHO and HOW property in our model.

Our model performs entity recognition by labelling terms, similar to the MEANT metric, and incorporates pretrained masked language models (BERT and its derivative models) to capture the sound descriptors of captions. Furthermore, similar to the SPIDEr-FL score, our metric integrates the fluency penalty from the FENSE metric. In summary, our approach involves computing the cosine similarity of the embeddings, focusing on the correspondence of sound descriptor categories. We also incorporate in our approach a fluency penalty and, in the absence of overlap, default to sentence-BERT embeddings.
 
\section{Methods}

The backbone of the proposed metric is a NLP model that is capable of classifying words from human (or model) generated captions into a set of sound descriptor categories. These categories reflect semantic attributes of the sound-generating objects and events that listeners derive from sounds, and have been derived from a recent survey of taxonomies of everyday sounds \cite{giordanoWhatWeMean2022a}  (see Table \ref{tab:lab-cat}). This word model is then applied to both the candidate and the reference captions, and a score is calculated that reflects the similarity of their sound descriptor categories.

\begin{table}[]
\centering
\caption{Sound descriptor categories used in calculating correspondence between captions.\\ Items marked with * exist only in the dataset of 13 labels.}
\label{tab:lab-cat}
\begin{tabular}{@{}ll@{}}
\toprule
Label                  & Description                           \\ \midrule
WHO                    & sound-generating agent                \\
WHO/WHAT PROPERTY*     & describes object or person            \\
WHAT                   & vibrating object or substance        \\
HOW                    & sound-generating actions/mechanisms                 \\
HOW PROPERTY*          & specifies action                      \\
WHEN                   & temporal context                      \\
WHERE                  & spatial context                       \\
WHAT/WHERE*            & objects that contribute to acoustics \\
SOUND TYPE             & sound-signal categories                \\
SOUND PROPERTY         & acoustic/auditory sound properties  \\
NON-AUDITORY SENSATION & non-auditory attributes of sound      \\
OTHER                  & labels that do not describe sound     \\
O                      & omitted labels                \\ \bottomrule
\end{tabular}
\end{table}

\subsection{Model training}

To obtain this word-classification model, a random subset of captions from the Clotho dataset was annotated. The Clotho dataset is widely used in AAC and consists of 29,645 captions for 5,929 audio excerpts, each 15 to 30 seconds long \cite{drossosClothoAudioCaptioning2019a}. Various pre-trained NLP models were fine-tuned to this dataset, and their performance compared to determine the most effective classifier (see below). Specifically, two word-labeled caption datasets were generated using the Prodigy web annotation tool \cite{ProdigyProdigyAnnotation}. In the initial training iteration, every word in each caption was labeled with one category from a set of 10 (dataset 1). For the final model, a set of 13 labels was used (dataset 2) (see Table \ref{tab:lab-cat}).

Dataset 1 consists of 2300 captions labeled by 3 annotators, with 1120 captions being duplicates across annotators for the estimation of annotator agreement statistics. The annotators labeled 694, 1387 and 219 labels captions, respectively. The average inter-annotator agreements (Cohen's kappa coefficient), averaged across annotator pairs, and annotations, were 0.808, 0.836 and 0.839.  To train our models, duplicates were filtered out to ensure that they did not lead to biased results from training items multiple times per epoch or data leakage in the test set, leaving 1158 captions in the final dataset. Of these 1158 captions, the annotators labeled 285, 727 and 146 respectively. 

We applied a similar procedure to dataset 2, which was labeled by two annotators, with a Cohen's kappa coefficient of 0.794. Dataset 2 is a subset of dataset 1, but contains more descriptor categories. Dataset 2 initially contained 989 captions, where the annotators covered 494 and 495 captions, respectively. After removing duplicates, the final dataset contained 500 unique captions. Although in dataset 2 there were more sound descriptors and less training data than dataset 1, this resulted in an increase in classification F1 scores. 

Using the 2 labeled datasets, various pre-trained Transformer encoder models were fine-tuned with a classification head (see Table \ref{tab:bert-models}), as implemented in the HuggingFace Transformers library \cite{wolfTransformersStateoftheArtNatural2020a}. During tokenization, each caption was split into tokens, each corresponding to a sound descriptor. Tokens from words that did not correspond to a sound descriptor in our dataset were assigned the label O and omitted. For training, we used an AdamW optimizer \cite{loshchilovDecoupledWeightDecay2019}, with a learning rate of 2e-5 and weight decay. The labeled dataset was divided into 80\% train and 20\% test sets. The model was trained for 5 epochs, since initial tests showed that after 5 epochs overfitting might occur, as identified by observing a growing difference between the training and validation losses after the fifth epoch. For each trained NLP model, the averaged and category-specific (Who-What-How-Where) F1-scores on the test set for the two distinct variants of models were obtained using 10 (top) or 13 (bottom) labels (Table \ref{tab:bert-models}). This F1-score gives an indication how well the model is capable of classifying the sound descriptors on unseen sentences. 

\begin{table}[]
\caption{Comparison of BERT-based models on token classification of our dataset. The highest scores are in bold. The groups of models are 10 labels (top) and 13 labels (bottom).}
\label{tab:bert-models}
\begin{tabular}{llllll}
\hline
Name        & F1             & F1 How         & F1 What        & F1 Where       & F1 Who         \\ \hline
\multicolumn{6}{c}{\cellcolor[HTML]{C0C0C0}\textit{\textbf{10 Labels}}}                          \\
BERT        & \textbf{81.6} & 88.3          & 80.6          & 78.5          & \textbf{85.3} \\
RoBERTa     & 81.2          & \textbf{89.5} & 79.3          & \textbf{79.2} & 80.0          \\
XLM RoBERTa & 79.7          & 85.8          & 80.5          & 78.0          & 81.0          \\
ALBERT      & 80.6          & 87.0          & \textbf{81.2} & 76.8          & 83.7          \\
DeBERTa     & 79.7          & 85.6          & 80.1          & 75.6          & 83.9          \\
\multicolumn{6}{c}{\cellcolor[HTML]{C0C0C0}\textit{\textbf{13 Labels}}}                          \\
BERT        & 80.3          & 90.8          & 80.6          & 86.0          & 91.2          \\
RoBERTa     & \textbf{84.2} & 91.8          & 84.7          & \textbf{92.3} & 90.9          \\
XLM RoBERTa & \textbf{84.2} & \textbf{93.7} & 85.6          & 83.9          & 92.9          \\
ALBERT      & 83.0          & 91.4          & \textbf{87.5} & 84.5          & \textbf{95.5} \\
DeBERTa     & 80.9          & 88.7          & 87.1          & 84.7          & 87.9          \\ \hline
\end{tabular}
\end{table}
 
\subsection{Metric definition}

The fine-tuned model is then applied to both the candidate and the reference captions. For example, for the sentence ``a person is walking on a hard surface'', groups of tokens \textit{person}, \textit{walking on} and \textit{hard surface} are categorized into \textit{WHO}, \textit{HOW} and \textit{WHAT/WHERE} respectively. In the case of 5 reference captions instead of 1 reference caption, the words per category in each of the reference captions are combined (e.g. WHO in ``bird caws''  ``bird croaks'' becomes ``bird bird'').  

For each of these group of words, a corresponding sentence embedding is calculated with SBERT \cite{reimers-2019-sentence-bert}. The cosine similarity is computed on overlapping sound descriptors (e.g. WHO in candidate and reference). This ensures focus on entities present in the candidate text and optimizes for better wording per descriptor category. However, a drawback is that it may neglect entities found in reference texts that do not exist in the candidate (see Appendix \ref{app:in-depth} for examples).

The cosine similarity of the overlap is calculated as follows. For each candidate and reference caption $C$ and $R$, for each candidate and reference token $c \in C$ and $r \in R$, given the sound descriptor of $c, r$: $\mathbbm{c}, \mathbbm{r} \in \mathbb{C}_l \cap \mathbb{R}_l$, the cosine similarity is calculated as:
\[
\text{CosSim}(c, r) = \frac{E(c)^\intercal \cdot E(r)}{\|E(c)\|\|E(r)\|}
\]
with $E(\cdot)$ representing the embedding of the token. As we use pre-normalized vectors, cosine similarity is reduced to $E(c)^\intercal \cdot E(r)$.
This cosine similarity score indicates the relation of each pair of tokens so that similar tokens in sound descriptors also get a high score, contrary to using the n-gram. 

From this cosine similarity, the maximum score of the precision (Pr) and recall (Re) overlap of the candidate and reference tokens is taken, similarly to BertScore \cite{zhangBERTScoreEvaluatingText2020}. 
\[
Pr(\mathbbm{c},\mathbbm{r}) = \frac{1}{|\mathbbm{r}|} \sum_{r \in \mathbbm{r}} \max_{c \in \mathbbm{c}} \text{ CosSim(}c,r\text{)}
\]
\[
Re(\mathbbm{c},\mathbbm{r}) = \frac{1}{|\mathbbm{c}|} \sum_{c \in \mathbbm{c}} \max_{r \in \mathbbm{r}} \text{ CosSim(}c,r\text{)}
\]
The F-score is computed by weighting 9 times more the recall of the token than the precision, as done similarly for the METEOR metric, and shown to improve the correlation with human judgement \cite{banerjeeMETEORAutomaticMetric2005}. Consistently, we also observed a positive correlation between emphasis on recall and better outcome on the FENSE benchmark.

The ACES metric puts focus on the token with the highest similarity score between the candidate and the references due to the maximization in the above formula. This approach has limitations. For instance, it may overlook irrelevant tokens even if they are categorized similarly. Additionally, the metric does not consider captions composed of compound sentences. In such cases, the ACES score emphasizes only on the category in one part of the compound sentence with the greatest similarity, disregarding the rest.

\[
F_{\text{score}}(\mathbbm{c},\mathbbm{r}) = \frac{10PrRe}{Re + 9Pr}
\]
All embeddings are then averaged: 
\[ 
F_{\text{single}}(C, R) = \frac{1}{|C \cap R|} \sum_{\mathbbm{c}, \mathbbm{r} \in C \cap R} F_{\text{score}}(\mathbbm{c}, \mathbbm{r})
\] 
If there is no overlap at all, we set the score to 0. 
As a last step, a small penalty is added to our scoring method. This penalty is designed to slightly decrease the scores of shorter sentences that include fewer entities. The penalty formula is as follows:
\[
\text{penalty} = ( |\mathbb{L}| - |\mathbb{C}_l \cap \mathbb{R}_l| ) \frac{1}{|\mathbb{L}|} \frac{1}{1850}
\]
$\mathbb{L}$ is defined by the total number of possible sound descriptors, in this case 13. $|\mathbb{C}_l \cap \mathbb{R}_l|$ is the number of sound descriptors that are the same in both the candidate and reference captions. This penalty is subtracted from the F-score for each pair of candidate and reference sentences, resulting in the following formula:
\[
\text{ACES}_{\text{1}}(C,R) = F_{\text{single}}(C,R) - \text{penalty}
\]
The reason for this penalty of 1850 is based on our belief that longer sentences, which usually have more entities, should be encouraged in AAC. However, tests showed that higher penalties did not lead to better performance on the FENSE benchmark. By applying only a slight penalty, a preference for detailed sentences is still included, and the metric still correlates well with human evaluation.

For each candidate and reference caption $C$ and $R$ a single value is returned. An extra fluency error detection penalty is added, similar to the one in the FENSE and SPIDER-FL metrics \cite{zhouCanAudioCaptions2022}. The best results were found by putting a weight of 0.5 instead of 0.9, the default value in the FENSE metric (see Appendix \ref{app:hyperparameters}). For every single candidate and reference caption, ACES is computed as: 
\[
\text{ACES}(C,R) = \text{ACES}_{\text{1}}(C,R) - 0.5 \times \text{fluency}(C) \times \text{ACES}_{\text{1}}(C,R)
\]

A combined ACES score for a whole dataset is calculated by taking the average across the pairs of candidate and reference captions.

\subsection{Hyperparameter tuning}
To find the optimal parameters for our model, a hyperparameter tuning phase is performed.  During hyperparameter tuning, the library WandB and its sweep function with Bayesian optimization \cite{wandb} carry out a search through 18 different parameters (see Appendix \ref{app:hyperparameters}, for an overview). The Bayesian optimization narrows the search scope by focusing on parameters showing the most promise. When the parameters are adjusted, the total score on Clotho-Eval is maximized, with the aim of outperforming the other models on the FENSE benchmark (see Table \ref{tab:metric-perf}). The emphasis for maximization is put on the Clotho-Eval benchmark, since its captions have longer sentences than Audiocaps-Eval (11.334 vs 8.796 words on average) and contain on average more entities (5.433 vs. 4.742). However, the initial results showed that optimizing for one benchmark also resulted in better results for the other benchmark.

The final parameters are from a run of a model that specifically outperforms some values on the FENSE benchmark and has a good overall result. Several configurations of the ACES model outperformed other metrics on the benchmark, but the final configuration also included values that resembled previous metric research on the METEOR and Fluency detection methods \cite{banerjeeMETEORAutomaticMetric2005, zhouCanAudioCaptions2022}. 

\subsection{Optimization}

Since the use case for this benchmark is during model validation, an essential part of the work is to optimize the model for implementation speed. The final model requires about ~6.4GB of RAM for the token classification part, and the model for the embeddings uses about 1.2 GB of RAM. This is optimized by quantizing the model weights to float16 precision, resulting in a model size reduction of about 50\%. Also, the model can easily perform distributed calculations by automatically scaling up to multi-GPU inference with the Accelerate Python package \cite{accelerate}.

Additionally, by relying on a smaller version of the RoBERTa model (RoBERTa-base instead of RoBERTa-large), we are able to achieve a 25\% speedup in the computation of ACES at the cost of a negligible drop in performance.

\section{Experiments}

We evaluated our models by comparing them to similar metrics in AAC.  In parallel, our models are tested against the FENSE benchmark to allow the estimation of its predictions of human evaluations. The FENSE benchmark consists of two datasets, Clotho-Eval and AudioCaps-Eval. These are derived from the Clotho and AudioCaps dataset, respectively, and contain human evaluations for 4 categories of pairs (see section \ref{sec:human-eval}).

\subsection{Model evaluation}

\begin{table}[]
\centering
\caption{Models were evaluated considering several metrics, including our proposed metric ACES.}
\label{tab:model-measuring}
\begin{tabular}{@{}llllll@{}}
\toprule
              & BERTScore & FENSE & ROUGE$_L$ & SPIDEr & ACES  \\ \midrule
\href{https://github.com/audio-captioning/dcase-2021-baseline}{Baseline 2021}  & 82.3     & 1.9 & 27.3    & 6.4  & 15.7 \\
\href{https://github.com/wsntxxn/AudioCaption}{AudioCaption}  & 89.1     & 37.7 & 28.9    & 14.2  & 42.2 \\
\href{https://github.com/felixgontier/dcase-2022-baseline}{Baseline 2022}  & 90.5     & 45.7 & 36.5    & 23.1  & 46.1 \\
PANNs + BART  & 90.7     & 46.8 & 37.6    & 25.2  & 47.9 \\
PASST + BART  & \textbf{90.8}     & \textbf{48.3} & 37.8    & 26.0  & 48.8 \\
\href{https://github.com/felixgontier/dcase-2023-baseline}{Baseline 2023}  & 90.7     & \textbf{48.3} & \textbf{38.7}    & \textbf{27.0}  & \textbf{49.3} \\ \bottomrule
\end{tabular}
\end{table}

We evaluated several AAC models using the BERTScore, FENSE, SPIDEr and the proposed ACES metrics to compare the metrics against each other. The DCASE baselines for this year and last year were used as benchmarks for the metrics. The 2021 baseline model was an encoder-decoder architecture based on GRU's \cite{choLearningPhraseRepresentations2014}, whereas the 2022 model had a VGGish encoder and a BART-based decoder. In addition to the 2022 model baseline, two other versions were added. In these two versions, the VGGish encoder network was replaced by a PANNs \cite{kongPANNsLargeScalePretrained2020} and PaSST \cite{koutiniEfficientTrainingAudio2022a} encoder, respectively. The 2023 model baseline is also included, which uses a special pretrained CNN encoder based on PANNs. 

\begin{table*}[h]
\centering
\caption{Performance of various metrics on Audiocaps and Clotho (bold = top values).}
\label{tab:metric-perf}
\begin{tabular}{@{}l|llllllllll@{}}
\toprule
Metrics       & \multicolumn{5}{l}{AudioCaps-Eval}                                                                 & \multicolumn{5}{l}{Clotho-Eval}                                             \\ \midrule
              & HC            & HI            & HM            & MM            & \multicolumn{1}{l|}{Total}         & HC          & HI            & HM            & MM            & Total         \\ \midrule
SPIDEr        & 53.2          & 89.9          & 84.1          & 55.2          & \multicolumn{1}{l|}{65.4}          & 47.9        & 88.1          & 67.9          & 52.5          & 59.8          \\
BERTScore     & 60.6          & 97.6          & \textbf{92.9} & 65            & \multicolumn{1}{l|}{74.3}          & 57.1        & \textbf{95.5} & 70.3          & 61.3          & 67.5          \\
BLEURT        & \textbf{77.3} & 93.9          & 88.7          & 72.4          & \multicolumn{1}{l|}{79.3}          & 59          & 93.9          & 75.4          & 67.4          & 71.6          \\
Sentence-BERT & 64            & \textbf{99.2} & 92.5          & 73.6          & \multicolumn{1}{l|}{79.6}          & 60          & \textbf{95.5} & 75.9          & 66.9          & 71.8          \\
FENSE         & 64.5          & 98.4          & 91.6          & \textbf{84.6} & \multicolumn{1}{l|}{\textbf{85.3}} & 60.5        & 94.7          & 80.2          & \textbf{72.8} & \textbf{75.7} \\ \midrule
SPICE+        & 59.1          & 85.4          & 83.7          & 49            & \multicolumn{1}{l|}{62}            & 46.7        & 88.1          & 70.3          & 48.7          & 57.8          \\
SPICE+emb     & 63.5          & 96.4          & 91.6          & 70            & \multicolumn{1}{l|}{77}            & \textbf{61} & 94.7          & 76.3          & 61.6          & 68.9          \\ \midrule
ACES          & 64.5          & 95.1          & 89.5          & 82.0          & \multicolumn{1}{l|}{83.0}          & 56.7        & \textbf{95.5} & \textbf{82.8} & 69.9          & 74.0         
\end{tabular}
\end{table*}

Table \ref{tab:model-measuring} displays the evaluation results of AAC models. ACES has more variation to BERTScore, which is helpful for discriminating results that are close to each other. ACES scores are consistent with other metrics. Notably, BERTScore and FENSE rate PASST + BART similarly to the Baseline 2023 model. Other metrics favor the baseline 2023 model, including ACES. This could indicate that ACES is a more reliable metric than BERTScore and FENSE, although further investigation is required.   

\subsection{Human evaluation} \label{sec:human-eval}

We used the FENSE benchmark \cite{zhouCanAudioCaptions2022} to evaluate our score against human evaluations, and to compare ACES to several other metrics (see Table \ref{tab:metric-perf}). The FENSE benchmark comprises four components, each designed to calculate the quality of a candidate metric compared to its reference: (i) human-correct (HC), (ii) human-incorrect (HI), (iii) human-machine (HM) and (iv) machine-machine  (MM). For each of these components, 4 human evaluators compare two candidate captions for each sound, and indicate which target sound is better (final outcome based on majority vote with target sound not considered in the absence of a majority). For the HC component, both candidate captions are generated by human annotators describing the same sound (human captioning data from Clotho and AudioCaps datasets) For HI, a caption generated by a human annotator for a target sound is compared with a human-generated caption for a different sound. The HM component compares instead a human caption with a caption for the same target sound generated with a captioning model. The MM component finally compares machine-generated captions from two captioning models (see \cite{zhouCanAudioCaptions2022}, for details). Audio captioning metrics are finally bench marked against the majority-based human evaluation of the candidate captions, with the final score measuring the proportion of target sounds for which the captioning metric and human evaluators agreed in their choice of the best captions among the two candidates.

On this FENSE benchmark, approaches that use computational similarity of contextualized embeddings (ACES, FENSE, BERTScore, BLEURT, Sentence-BERT, SPICE+emb), are favored over approaches that use entity labeling (SPIDEr, SPICE+). This could indicate that participants’ judgments of descriptions from the FENSE benchmark are correlated with distances between descriptions in the BERT embedding space.

\section{Discussion}
Overall, the ACES metric demonstrates comparable performance to other metrics in the FENSE benchmark, and notably outperforms all other metrics in the human-incorrect (HI) and human-machine (HM) categories on the Clotho-Eval dataset. ACES provides a versatile backbone that can be used to recognize a sentence's entities, which can be helpful for explainability purposes.

ACES calculates a score based on semantic descriptors and shows promising results in comparison to other metrics on the FENSE benchmark. A comparison between metrics with several examples is shown in Appendix \ref{app:in-depth}. This appendix also showcases some drawbacks of our approach, including cases where there is no overlap of semantic descriptors. In case of no overlap, ACES assigns the score 0. It also only takes the most similar entity into account: in case of lower sentences with linking words, it focuses on the most similar word per semantic descriptor. 

Results from candidate captions that exhibit an higher ACES score contain several sound descriptors that are high in cosine similarity of the vector representation to the ground truth vector representation. These sound descriptors are important to include in a caption, as they represent listeners' verbal descriptions of everyday sounds.  

\section{Conclusion}

We introduced Audio Captioning Evaluation on Semantics of Sound (ACES), a novel AAC evaluation metric based on audio semantics. ACES combines both semantic similarities and semantic entity labeling, resulting in an approach for evaluating audio captions that departs from earlier work relying exclusively on embeddings, or n-grams, or entities. ACES combines elements from earlier research on AAC evaluation and outperforms other metrics on the FENSE benchmark. 

Future research can explore several promising directions, such as the involvement of large language models such as GPT-4 to evaluate sentences for fluency issues and to grade semantic richness and factual correctness in generated captions. A strategy for this could be to train a large language model on how a human would evaluate a dataset of generated audio captions and create an automatic pipeline for the caption evaluation.

\section*{Acknowledgment}
\addcontentsline{toc}{section}{Acknowledgment}
This work was supported by the Dutch Research Council (NWO 406.20.GO.030 to Prof. Elia Formisano), the French National Research Agency (ANR‐21‐CE37‐0027‐01 to Bruno L. Giordano; ANR-16-CONV-0002 -- ILCB; ANR11-LABX-0036 -- BLRI),
Data Science Research Infrastructure (DSRI; Maastricht University) and the Dutch Province of Limburg.

\bibliographystyle{IEEEtranN}
\bibliography{library.bib}

\appendices

\section{Overview of hyperparameters}

\noindent In bold the parameter that performed the best in the hyperparameter tuning. Percentages are correlation scores with Clotho-Eval Total in total for all runs. See also the \href{https://api.wandb.ai/links/aud2sem/md9xlq8h}{WandB Report}.
\label{app:hyperparameters}
\begin{itemize}
\item \textit{model}: ("gijs/aces-roberta-10", \textbf{"gijs/aces-roberta-13"}) Selects which pre-trained ACES model to use for named entity recognition and embedding extraction. The options are the 10 label or 13 label versions, which detect different numbers of entity types.

\item \textit{fl\_weighing}: (\textbf{True}, False) (34.9\%). Determines whether a fluency model should be added to the model. This model detects sentences with errors in them and penalizes them with \textit{F1\_weight} (float between 0.0 and 1.0) (1.4\%): which controls the amount the score is reduced if an error is detected.

\item \textit{average\_strategy}: ("simple", "first", \textbf{"max"}, "average") Controls how the per-token predictions from the ACES model are aggregated into entities.

\item \textit{use\_sbert}: (\textbf{True}, False) (22.7\%) Determines if SentenceBERT is used to calculate entity embedding similarity rather than the native ACES embeddings.

\item \textit{division}: (float between 0.0 and 1.0. \textbf{0.998}) (-2.6\%) Controls weighting between the cosine similarity score and the entity overlap score.

\item \textit{F1\_calc}: ("mean", "mean-max", \textbf{"max-mean"}, "max-max"). Sets how precision and recall are combined to calculate F1 between the candidate and reference entities.

\item \textit{F1\_beta}: (float between 0.1 and 10.0. \textbf{9.0}) (1.9\%) Controls beta parameter that balances precision and recall for F-beta score for the cosine similarity

\item \textit{F1}: (float between 0.1 and 10.0. \textbf{3.798}) (9.3\%) Controls the beta parameter that balances precision and recall for F-beta score for the overlap.

\item \textit{apply\_penalty}: (\textbf{True}, False) (-8.6\%). Boolean for whether to apply the entity overlap penalty. This is to penalize sentences that have only a few classes, as we prefer longer more descriptive sentences over shorter ones (``bird caws noisily with an alarming tone'' over just ``bird caws'').  \textit{penalty\_score} (int between 1 and 2000. \textbf{1850}) (-5.5\%): Penalty applied when there is no entity overlap. Higher values decrease this penalty. If there are a lot of labels in the sentences, it still corresponds to a high score.

\item \textit{overlap\_type}: ("cand", "ref", \textbf{"both"}, "F1"). Sets the formulation of entity overlap used. Based on the candidate, reference, both, or F1-weighted overlap.

\item \textit{distance\_technique} (\textbf{"cosine"}, "Euclidean"): Distance metric to calculate entity embeddings.

\item \textit{use\_score}: ("pairwise", "mean", \textbf{"no"}). With the pairwise score, both the combination of the similarity score and the overlap of the entities per combination of sentences are calculated. "mean" corresponds to calculating the overlap after taking the mean. \textit{score\_weighing} (float between 0.0 and 1.0. \textbf{0.5}) (18.0\%): Controls balance between embedding similarity and overlap when combining values.

\item \textit{overall\_sbert}: (True, \textbf{False}). Whether to incorporate an additional overall SentenceBERT similarity between the candidate and reference. \textit{overall\_sbert\_weight} (float between 0.0 and 1.0. \textbf{0.5}) (-8.2\%): Weighting on overall SBERT similarity when used. \textit{sbert\_based\_on\_scores} (\textbf{True}, False) (4.3\%): Only uses SBERT similarity if the ACES score is 0, which is the case when there are no labels at all present.

\end{itemize}

\section{In depth comparison of ACES with previous metrics} \label{app:in-depth}
This Appendix presents selected examples of the FENSE benchmark, featuring both successful and unsuccessful cases of scoring from the ACES metric. The examples are displayed in Table \ref{tab:metric-examples}. The first three examples demonstrate situations where ACES outperforms FENSE and other measures. The final two instances reveal ACES's limitations. 

In the initial example, ``a door is followed by a'',  both SPICE and ACES score 0.0 due to their reliance on overlapping semantic pairs. SPICE utilizes nonexisting graph overlap based on its semantic graph, and ACES relies on its semantic sound categories in the captions. For ACES it returns 0 due to ``door'' being the sole recognized WHAT entity and in the reference captions no WHAT entities. Other metrics, however, return similar, albeit higher, scores. 

The second example highlights a more pronounced gap between the ACES and FENSE scores. The lower score for ACES arises because of the difference in length between the references and candidate captions, affecting the overlap in categories considered. The candidate caption is classified into three categories. WHO (``birds'', ``young peoples''), HOW: (``cackling''). WHAT: (``voices''). For WHAT (``voices''), ACES matches it to the only WHAT in the references: ``wings'', which drives the score down. However, less descriptive sentences may be the origin of the lower score. 

For the third scenario, ACES again assigns a lower score compared to other metrics. This is due to the reference caption containing only WHO and HOW categories, creating a mismatch on HOW between the candidate and references. The candidate's ``rolling'' and ``strikes'' differ from the reference's ``roars'', ``falls'' and ``falling''. Despite similar overall overlap, they are missing some key components, i.e. no mention of lightning in any of the references. 

The fourth example exhibits considerable overlaps resulting in a higher ACES score. However, certain elements, such as ``in a quiet environment'', are absent. In the candidate caption, there are no similar WHERE matches, which results in ACES wrongly attributing a high score to a caption missing contextual details.

Finally, the fifth example highlights another instance where ACES scores higher on average compared to other metrics. This results from the similarity in WHO and HOW categories between the candidate and reference captions. Yet, ACES does not fully discern the difference between ``speaking'' and ``shouting'', focusing instead on the entity with the greatest overlap, in this case, the HOWs ``squeaking'' and ``bouncing'', which are present in both the candidate and reference captions.

\begin{table*}[]
\centering
\caption{Candidate and reference examples from the FENSE benchmark with corresponding AAC metrics}
\label{tab:metric-examples}
\begin{tabular}{@{}llllllll@{}}
\toprule
Candidate                                                                                   & References                                                                                                                                                                                                                                                                                                                                                                                                                                                                                 & METEOR & ROUGE$_L$ & SPICE  & BERTScore & FENSE  & ACES   \\ \midrule
A door is followed  by a                           & \begin{tabular}[c]{@{}l@{}}Some scratching and rustling \\ with small clicks\\ \\ Rustling and some knocks\\ \\ Some rustling and scratching \\ with a short click\\ \\ Continuous clanking and rustling\end{tabular}                                                                                                                                                                                                                                                                      & 0.0265 & 0.1393  & 0.0    & 0.9814    & 0.0311 & 0.0    \\ \midrule
\begin{tabular}[c]{@{}l@{}}Birds cackling and \\ young peoples voices\end{tabular}          & \begin{tabular}[c]{@{}l@{}}A man speaking followed by \\ pigeons cooing and flapping wings \\ then a kid speaking and someone \\ claps loudly\\ \\ A man talking then a young boy \\ talking followed by a loud pop\\ as pigeons coo and bird wings flap\\ \\ A man talking followed by pigeons\\ cooing and bird wings flapping\\ then a young man talking\\ \\ A male voice speaks and a bird coos\\ and flaps its wings\end{tabular}                                                    & 0.0440 & 0.1680  & 0.0541 & 0.9646    & 0.6621 & 0.3230 \\ \midrule
\begin{tabular}[c]{@{}l@{}}Rolling thunder with \\ lightning strikes\end{tabular}           & \begin{tabular}[c]{@{}l@{}}Thunder and a gentle rain\\ \\ Thunder roars in the distance as \\ rain falls\\ \\ Rain falling with thunder in the \\ distance\\ \\ Rain and thunder\end{tabular}                                                                                                                                                                                                                                                                                              & 0.1741 & 0.2618  & 0.2000 & 0.9858    & 0.6899 & 0.4755 \\ \midrule
\begin{tabular}[c]{@{}l@{}}Typing on a computer\\ keyboard\end{tabular}                     & \begin{tabular}[c]{@{}l@{}}Typing on a keyboard is occurring\\ in a quiet environment\\ \\ Typing on a keyboard is ongoing\\ in a quiet environment\\ \\ Typing on a keyboard is occurring\\ in a quiet environment\\ \\ Typing on a keyboard is ongoing\\ in a quiet environment\end{tabular}                                                                                                                                                                                             & 0.2049 & 0.5031  & 0.2500 & 0.9882    & 0.7101 & 0.9677 \\ \midrule
\begin{tabular}[c]{@{}l@{}}Squeaking and bouncing\\ followed by a man speaking\end{tabular} & \begin{tabular}[c]{@{}l@{}}Several basketballs bouncing and \\ shoes squeaking on a hardwood \\ surface as a man yells in the distance\\ \\ A man yelling in the background as \\ several basketballs bounce and shoes\\ squeak on a hardwood surface\\ \\ A man yelling in the distance as \\ several basketballs bounce and shoes\\ squeak on hardwood floors\\ \\ Multiple basketballs bouncing on a \\ hard surface and shoes squeaking\\ as a man shouts in the distance\end{tabular} & 0.1373 & 0.2254  & 0.0    & 0.9752    & 0.5331 & 0.8676 \\ \bottomrule
\end{tabular}
\end{table*}

\end{document}